\documentstyle[12pt]{article}
\def\dspace{\baselineskip = .30in}
\def\strut{\rule[-.5cm]{0cm}{1cm}}
\begin{document}

\title{ MSSM and Large $tan\beta$ from SUSY Trinification}

\author{{\bf G. Lazarides and C. Panagiotakopoulos}\\ Physics Division\\
School of Technology\\ University of Thessaloniki\\
Thessaloniki, Greece}

\date{ }
\maketitle

\dspace
\centerline{\bf Abstract}
\vspace{.2in}
We construct a supersymmetric model based on the semi-simple gauge
group $SU(3)_c \times SU(3)_L \times SU(3)_R$ with the
relation $tan\beta \simeq m_t/m_b$ automatically arising from its
structure. The model below a scale $\sim 10^{16}$ GeV gives naturally
rise just to the minimal supersymmetric standard model and therefore
to the presently favored  values for $sin^2 \theta_w$ and $\alpha_s$
without fields in representations higher than the fundamental.

\newpage
\par
The remarkable observation$^{(1)}$ that the renormalization group equations of
the minimal supersymmetric standard model (MSSM) with a supersymmetry
(SUSY) breaking scale $M_s \sim 1$ TeV are astonishingly consistent
with the observed values of $sin^2 \theta_w$ and $\alpha_s$ and
unification of the three gauge couplings at a scale $M_X \sim 10^{16}$
GeV strongly suggests embedding of the standard gauge group $G_S \equiv
SU(3)_c \times SU(2)_L \times U(1)_Y$ in a larger one. It would
certainly be very desirable and perhaps a guide to  the correct
extention if this embedding led to the determination of the important
MSSM parameter $tan\beta$. This parameter is defined as the ratio of the
vacuum expectation values (VEVs) of the doublet $h^{(1)}$ giving mass
to the up-type quarks and the doublet $h^{(2)}$ giving mass to the
down-type quarks and charged leptons. The embedding of the MSSM in the
minimal SUSY SU(5) model fails to determine $tan\beta$. However, SUSY
grand
unified theories (GUTs) based on larger groups, like SO(10), may
lead$^{(2)}$
to the asymptotic relation $tan\beta \simeq m_t/m_b$.

\par
We recently realized$^{(3)}$ that, for this relation to hold, one needs much
less than the enormous SO(10) gauge  symmetry. It suffices the
left-right symmetric extention $G_{LR} \equiv SU(3)_c \times SU(2)_L
\times SU(2)_R \times U(1)_{B-L}$ of the standard gauge group $G_S$
and the further assumption that the higgs doublets $h^{(1)}, h^{(2)}$
as well as the 3rd generation $SU(2)_L$ -singlet antiquarks  $u^c,
d^c$ form two $SU(2)_R$ -doublets $h$ and $q^c$ respectively. Then,
the 3rd generation quark mass terms originate from the single $G_{LR}$
-invariant term $fqq^ch$  ($q$ is the $SU(2)_L$ - doublet quark)
with the  unique yukawa coupling $f$ and the relation $tan\beta = m_t/m_b$
follows immediately.
Under analogous assumptions for the leptons, we further obtain the
relation $tan\beta = m^D_{\nu_{\tau}} /m_\tau$ ($m^D_{\nu_{\tau}}$
being the Dirac mass of the $\tau$ -neutrino). We then formulated
widely applicable sufficient conditions on SUSY GUTs  guaranteeing
that the situation is  as described above. Thus, we opened up the
way towards constructing models based on simple as well as on
semi-simple groups with $tan\beta \simeq m_t/m_b$.

\par
SUSY GUTs based on semi-simple gauge groups are very interesting for
various reasons$^{(4)}$ the most important one being related to the proton
decay problem. It is well-known that in minimal SUSY GUTs there is a
close relationship between light quark masses and proton decay
amplitude mediated by color triplet and antitriplet higgsino exchange.
This relationship makes it difficult to forbid proton decay by
imposing discrete symmetries without eliminating the light quark mass
terms. In contrast, this is easily achieved in SUSY GUTs based on
semi-simple groups. Another property making semi-simple groups
attractive is that they offer
the possibility  to construct viable
models using only fundamental representations and singlets. Also, with
semi-simple groups the gauge hierarchy problem might become easier to
solve. To  illustrate these nice features of semi-simple groups,  we
recently constructed$^{(5)}$ a SUSY GUT based on the  maximal subgroup $G
\equiv SU(3)_c \times SU(3)_L \times SU(3)_R$ of $E_6$ using only
fields contained in 27, $\overline{27}$ and singlet representations of
$E_6$ in which, below a scale $\sim 10^{16}$ GeV, we naturally recover
the MSSM.

\par
The models with $tan\beta \simeq m_t/m_b$ constructed in ref.(3) either
make use of adjoint higgs fields or do not give rise, below a scale
$\sim 10^{16}$ GeV, just to the MSSM with the successful $sin^2
\theta_w$ and $\alpha_s$ predictions following from unification of
the three $G_S$ couplings at this scale. The purpose of the present
paper is to naturally derive for the first time the MSSM with the
usual predictions for $sin ^2 \theta_w$ and $\alpha_s$ from a SUSY
GUT without adjoint higgs fields and with the relation $tan\beta
\simeq m_t/m_b$ being automatically guaranteed.

\par
Our model based on the gauge group $G \equiv SU(3)_c \times SU(3)_L \times
SU(3)_R$
 is  a variation of the model of ref.(5)  retaining all its good
features.
We assume that the theory emerges as an effective theory
from a more fundamental theory at a scale $M_c = M_P/ \sqrt{8 \pi}$
close to the Planck
mass $M_P \equiv 1.2 \times 10^{19}$ GeV. At the scale $M_c$ the three
$SU(3)$ gauge couplings are equal and, due to a symmetric spectrum among
them, they remain equal (to one loop)
until a unification  scale $M_X \sim 10^{16}$ GeV where $G$ breaks down to
$G_S$. Below the scale $M_X$ our model gives naturally rise exactly to
the MSSM.
Using appropriate discrete symmetries we forbid proton decay, we
solve the gauge hierarchy problem and we ensure that the relation
$tan\beta \simeq m_t/m_b$ holds. Only fields contained in the
27-dimensional and singlet representations of $E_6$ are used and
generic superpotential couplings are assumed throughout.

\par
The gauge non-singlet left-handed lepton, quark and antiquark
superfields transform under $G$ as follows:
\begin{displaymath}\begin{array}{lcccc}
\lambda & = & (1, \bar{3}, 3) & = & \left(\begin{array}{ccc}H^{(1)} &
H^{(2)} & L\\ E^c & \nu^c & N\end{array}\right), \strut\\
Q & = & (3,3,1) & = &
\left(\begin{array}{c}q\\g\end{array}\right), \strut\\
Q^c & = & (\bar{3}, 1, \bar{3}) & = & \left(\begin{array}{c} u^c,
d^c,g^c\end{array}\right).\end{array}
\end{displaymath}
Here $N$ and $\nu^c$ denote $G_S$ -singlet superfields while  $g(g^c)$
is an additional down-type quark (antiquark). We will be working with
eight fields of each type $(\lambda, Q, Q^c)$ and five corresponding
mirror fields $(\bar{\lambda}, \bar{Q}, \bar{Q}^c)$. Notice that the
field content  ensures identical running of the three
$SU(3)$
gauge couplings to the one loop approximation in the $G$ -symmetric phase.

\par
We impose invariance under three $Z_2$ symmetries $P, C$ and $S_1$ and a
$Z_3$ symmetry $S_2$. Under $P$, all $\lambda, \bar{\lambda}$ fields
remain invariant while all $Q, \bar{Q}, Q^c, \bar{Q}^c$ fields change
sign. Under $C$, all fields remain invariant except $\lambda_6,
\lambda_8, \bar{\lambda}_3$ and $\bar{\lambda}_5$ which change sign.
Under $S_1$, all fields remain invariant except $\lambda_3, \lambda_6,
\lambda_7, \bar{\lambda}_3, \bar{\lambda}_4$, $Q_1, Q_2$ and $Q^c_3$
which change sign. Finally, under the generator of $S_2$, all fields
remain invariant except $\lambda_1, \lambda_2, Q_1, Q_2$,  $Q^c_3$
which are multiplied by $\alpha$ and $\lambda_3, Q_3, Q_1^c$, $Q^c_2$
which are
multiplied by $\alpha^2 (\alpha = e ^{2\pi i/3})$.

\par
The symmetry breaking of $G$ down to $G_S$ is obtained through
appropriate superpotential couplings of the fields acquiring a
superlarge VEV. These are the fields $\lambda_7, \bar{\lambda}_4,
\lambda_8, \bar{\lambda}_5$. The superlarge VEVs are: $<N_7> =
<\bar{N}_4>^* =$ $<\nu^c_8> = <\bar{\nu}^c_5>^*$. We also introduce in the
superpotential explicit mass terms of order $M_X$ (wherever allowed by
the discrete symmetries)
 for  $Q, \bar{Q}
$ and $Q^c, \bar {Q}^c$ pairs as well as the $\lambda, \bar{\lambda}$
pairs involving fields which do not acquire  superlarge VEVs.

\par
The above VEVs leave the discrete symmetries $P$ and $S_2$ unbroken. The
discrete symmetry $C$ combined with the diagonal $Z_2$ subgroup of the
 center of $SU(2)_L \times
SU(2)_R$ gives a $Z_2$ group $C^\prime$ which remains unbroken by all
VEVs and acts as "matter parity". The  discrete symmetry $S_1$
combined with the $Z_2$ subgroup of $SU(3)_R$ generated by the element
$diag (-1, +1, -1)$
gives a $Z_2$ group $S^\prime_1$ which remains
unbroken by the superlarge VEVs. The symmetry $P$ leads to a practically
stable proton. The "matter parity" $C^\prime$ suppresses some lepton
number violating couplings at the level of the MSSM (which is
important for keeping the already generated baryon asymmetry of the
universe) and stabilizes the lightest supersymmetric particle. The
combined effect of $C^\prime$ and $S^\prime_1$ solves in a natural
manner the gauge hierarchy problem. Finally, the symmetry $S_2$ forces
the relation $tan\beta \simeq m_t/m_b \simeq m^D_{\nu_{\tau}}/m_\tau$.

\par
To see how these properties are realized,
we now turn to a brief discussion of the mass spectrum. All $G_S$
-non-singlet states which remain invariant under $S_2$ acquire masses
$\sim M_X$. The remaining $G_S$ -non-singlets are all contained in
$\lambda_1, \lambda_2, \lambda_3, Q_1, Q_2, Q_3, Q^c_1, Q^c_2, Q_3^c$
and transform non-trivially under $S_2$. These fields, because of the
unbroken $S_2$ symmetry, can only form mass terms by pairing up among
themselves. Also, because they all remain invariant under $C$, the only
$\lambda$ with large VEV that can be used in the renormalizable
superpotential to generate these mass terms is the $C$ -invariant
$\lambda_7$ which acquires a VEV in the $N$ direction and therefore
leaves the  $SU(2)_R$ symmetry unbroken. We conclude that all $G_S$
-non-singlet states which remain massless after taking into account
the above mass terms are necessarily either $SU(2)_R$ -singlets or
partners in $SU(2)_R $ -doublets. It is straightforward to see that
these states are the $q_i, u^c_i, d^c_i$, $L_i, E^c_i$
 with $i$=1,2,3,
one linear combination of $H^{(1)}_1$ and $H^{(1)}_2$,
which we call $h^{(1)}$, and its partner $h^{(2)}$ in the $SU(2)_R$
-doublet $h$ to which it belongs. These are exactly the states of the MSSM
and can, of course, be  classified by their charges under the unbroken
discrete symmetries $P, C^\prime, S^\prime_1$ and $S_2$. From the
$G_S$ -singlet states, we are only interested in the $C^\prime = - 1$
sector  because it contains  the right-handed neutrinos. In this
sector all states acquire masses $\sim M_X$ or $\sim M^2_X/M_c$ except
the $\nu^c_i$  which are the partners of $E^c_i$ in their
$SU(2)_R$ -doublets  $\ell^c_i$ and remain massless as expected.

\par
{}From the above discussion  we see that the electroweak higgs doublets
$h^{(1)}, h^{(2)}$ form a single $SU(2)_R$ -doublet $h$, the light
$SU(2)_L$ -singlet antiquarks $u^c_i, d^c_i$ form three  $SU(2)_R$
-doublets $q^c_i$ and  the light $SU(2)_L$ -doublet quarks $q_i$ are
$SU(2)_R$ -singlets. In the lepton sector, the $SU(2)_L$
-singlet charged antileptons belong to
three $SU(2)_R$ - doublets $\ell^c_i$ and
the $SU(2)_L$ -doublet leptons $L_i$ are $SU(2)_R$ -singlets.
Furthermore, the unbroken discrete symmetries allow  tree-level
masses at least for the 3rd generation. In the quark sector there is
only one such mass term allowed, namely $q^\prime q^c_3 h$ (where
$q^\prime$ is one linear combination of $q_1, q_2$). In the lepton
sector the allowed mass terms are $L_k \ell^c_m h $ with $k,m=1,2$. The
dominant among these mass terms
corresponds to the 3rd generation and contains the only allowed $\tau$
-neutrino Dirac mass term.
This suffices
to prove the relation $tan\beta =m_t/m_b = m^D_{\nu_{\tau}}/m_\tau$.
Analogous relations for the lighter quarks are not expected to hold.

\par
With the symmetry $S_2$ unbroken, Majorana masses for the states
$\nu^c_i$ are forbidden. Therefore $S_2$ has to break at a superheavy
scale. To this end, we introduce two gauge singlets $T_1$ and $T_2$
which remain invariant under $P, C$ and $S_1$ but get multiplied by
 $\alpha$ under
the generator of $S_2$. Let $T^\prime$ be the linear combination of
$T_1$ and $T_2$ which has a coupling of the type $T^\prime \lambda_3
\bar{\lambda}_4$ and $T$ the orthogonal linear combination. We assume
that $T$ acquires a VEV of the order of $M_X$ thereby breaking the $S_2$
symmetry. One can check that this breaking generates mixings among
states with different $S_2$ charges which are, however, at most $\sim
M_X/M_c$. Therefore our relations for $tan\beta$ become now only very
good approximations. This breaking leads to generation of Majorana
masses which are $\sim M^3_X/M^2_c$ for $\nu^c_k$ $(k=1,2)$ and $\sim
M^4_X/M^3_c$ for $\nu^c_3$. This, in turn, means that the $\tau$
-neutrino mass is close to 10 eV and therefore contributes significantly
to the hot dark matter of the universe. The other two neutrinos could
be light enough to allow for a solution of the solar neutrino puzzle
through the  MSW effect. The $\nu^c$ Majorana masses are also of the
correct order of magnitude to allow for the generation of lepton
number asymmetry$^{(4)}$ in the universe which at temperatures close to $M_W$
will be transformed into the observed baryon asymmetry.

\par
Finally, we point out that neither the $S^\prime_1$ symmetry can be
left unbroken because it forbids a higgsino mass term of the type
$h^{(1)} h^{(2)}$. The breaking of $S^\prime_1$ can be done using
superheavy VEVs acquired by gauge singlets $\tilde{T}$ which change
sign under $S^\prime_1$. If such singlets transform as $\beta=e^{2 \pi
i/N}$ or $\beta^{-1}$ under an additional $Z_N$ symmetry with
$N=odd$ then superpotential terms like $h^{(1)} h^{(2)} N_7
\frac{T}{M_c} (\frac{\tilde{T}}{M_c})^N$ give naturally rise to a
sufficiently small higgsino mass.

\par
We constructed a SUSY GUT based on the semi-simple group $SU(3)^3$
which below a scale $M_X \sim 10^{16}$ GeV gives naturally rise to the
MSSM and therefore to the presently favored values of $sin^2
\theta_w$ and $\alpha_s$. Besides, the relations $tan\beta \simeq
m_t/m_b \simeq m^D_{\nu_{\tau}}/m_\tau$ hold as automatic consequences
of the structure of the theory. This is the first example of a SUSY
GUT where these two features coexist and no use of fields in
representations higher than the fundamental is made.

\newpage
\section*{References}

\begin{enumerate}

\item S. Dimopoulos and H. Georgi, Nucl. Phys. \underline{B193} (1981)
150;
J. Ellis, S. Kelley and D. V. Nanopoulos, Phys. Lett. \underline
{B249} (1990) 441; U. Amaldi, W. de Boer and H. Furstenan, Phys. Lett.
\underline {B260} (1991) 447; P. Langacker and M. X. Luo, Phys. Rev.
\underline {D44} (1991) 817.

\item B. Anathanarayan, G. Lazarides and Q. Shafi, Phys. Rev.
\underline{D44} (1991) 1613; H. Arason, D. J. Castano, B. E.
Keszthelyi, S. Mikaelian, E. J. Piard, P. Ramond and B. D. Wright,
Phys. Rev. Lett. \underline{67} (1991) 2933; S. Kelley, J. L. Lopez and
D. V. Nanopoulos, Phys. Lett. \underline {B272} (1992) 387.

\item G. Lazarides and C. Panagiotakopoulos, Thessaloniki
Univ. preprint UT-STPD-1-94 (1994).

\item G. Lazarides, C. Panagiotakopoulos and Q. Shafi, Phys. Lett.
\underline{B315} (1993) 325 ; E. \underline{B317} (1993) 661.

\item G. Lazarides and C. Panagiotakopoulos, Thessaloniki Univ.
preprint UT-STPD-2-93 (1993).
\end{enumerate}
\end{document}